\documentstyle[12pt]{article}

\newcommand{\extraspace}{\addtolength{\abovedisplayskip}{2mm} 
                        \addtolength{\belowdisplayskip}{2mm} 
                        \addtolength{\abovedisplayshortskip}{2mm} 
                        \addtolength{\belowdisplayshortskip}{2mm}} 
\newcommand{\be}{\begin{equation}\extraspace} 
\newcommand{\ee}{\end{equation}} 
\newcommand{\bea}{\begin{eqnarray}\extraspace} 
\newcommand{\eea}{\end{eqnarray}} 
 
\newcommand{\nonu}{\nonumber \\[2mm]}

\newcommand{\half}{{\frac{1}{2}}}

\newcommand{\eps}{\epsilon}

\newcommand{\la}{\lambda}

\newcommand{\del}{\partial}

\begin{document} 
 
\vskip 2.5cm
\begin{center}

{\bf\LARGE  Non-abelian Exclusion Statistics}\\
\vskip 15mm
{\bf\large Sathya Guruswamy and Kareljan Schoutens}
\vskip 3mm  
{\sl Institute for Theoretical Physics \\
     Valckenierstraat 65, 1018 XE Amsterdam \\
     THE NETHERLANDS} 
\vskip 12mm
{\bf Abstract}
\end{center}
\baselineskip=15pt
\vskip 5mm
\begin{quote}
We introduce the notion of `order-$k$ non-abelian 
exclusion statistics'. We derive the associated
thermodynamic equations by employing the Thermodynamic 
Bethe Ansatz for specific non-diagonal scattering matrices. 
We make contact with results obtained by different 
methods and we point out connections with `fermionic 
sum formulas' for characters in a Conformal Field Theory.
As an application, we derive thermodynamic distribution
functions for quasi-holes over a class of non-abelian 
quantum Hall states recently proposed by Read and 
Rezayi.
\end{quote}

\vfill 


\vskip 4mm
\noindent Report number: ITFA-98-29
\hfill

\newpage
      
\section{Introduction}

It is well-known that strongly interacting quantum many-body systems in
low dimensions can have quasi-particles that are very different from the
microscopic degrees of freedom from which the system is built. A
spectacular phenomenon, which has been demonstrated in theory and
experiment, is that of `quantum number fractionalization': the quantum
numbers of the fundamental quasi-particles can be fractions of those of
the constituent degrees of freedom. Fractionally charged quasi-particles
over fractional quantum Hall (fqH) states are a proto-typical example. 
Along with unusual quantum numbers, quasi-particles in low-dimensional
systems can have equally unusual quantum statistics properties. In
many cases of interest, fundamental quasi-particles do not satisfy the
standard Pauli exclusion principle, but instead a generalization thereof. 

There are several angles from which unusual exclusion statistics for
quasi-particles in low dimensions may be studied. For systems that are
integrable by Bethe Ansatz, the systematics of solutions to the Bethe
equations can be interpreted in terms of exclusion statistics of the
`Bethe Ansatz quasi-particles' \cite{Ha1,3-potts}. A second approach
starts from scattering data of quasi-particles and associates statistics
properties via the Thermodynamic Bethe Ansatz (TBA) \cite{BW,FLS}. A third
approach \cite{Sc1,BS2} relies on algebraic properties of field operators
that describe quasi-particles in the context of a Conformal Field Theory
(CFT). There are various connections among the approaches mentioned here,
and in cases where they can be compared they lead to equivalent results.
[We refer to \cite{BS2} for further introduction and additional
references.]

Of particular interest are systems whose quasi-particles obey `fractional
exclusion statistics' as defined by F.D.M.~Haldane in 1991 \cite{Ha2}. 
Examples
are Calogero-Moser models of many-body quantum mechanics and the CFTs
describing edge excitations over abelian quantum Hall states \cite{ES}. In
the latter cases, the fractional exclusion statistics of edge
quasi-particles reflect the fractional (anyonic) braiding statistics of
bulk quasi-particles over the same quantum Hall state.

It is a well-known fact that the braid group for particles in two space
dimensions admits representations that have dimension higher than one.
This means that for a given number of particles with all positions and
quantum numbers fixed, more than one quantum mechanical state is possible.
On the corresponding state vectors, the braid group is represented by a
non-trivial matrix. Since matrices in general do no not commute, one
speaks of non-abelian statistics. In the context of the fractional quantum
Hall effect, explicit realizations of such non-abelian braid statistics
have been proposed \cite{MR, NW, RR, AS}. The bulk quasi-particles over
what are loosely called `non-abelian quantum Hall states', have
counterparts at the edge. In a recent publication \cite{Sc2}, one of us
employed the CFT approach of \cite{Sc1} to study the exclusion statistics
of edge quasi-holes over particular non-abelian quantum Hall states known
as the $q$-pfaffians. The resulting statistics are a proto-typical example
of what we call `non-abelian exclusion statistics'. 

In later work, other examples of `non-abelian exclusion statistics' have
been presented. These include quasi-particles over more general
non-abelian fqH states \cite{FNS}, spinons in spin $S\geq1$ Heisenberg
spin chains \cite{BLS2,FS}, generalized fermions in minimal models of CFT
\cite{BS2}, and spinons in some of the level-1 Wess-Zumino-Witten (WZW)  
models of CFT \cite{BS2}. In \cite{BS1} it was argued that the spinon 
(kink) quasi-particles in specific $SO(5)$-invariant 2-leg ladder models 
for strongly correlated electrons are non-abelions that carry the quantum 
numbers of physical electrons. 

What has been lacking until now is a unified description of the
thermodynamics associated to the various examples of non-abelian 
exclusion statistics. In this paper we write down an $S$-matrix for 
scattering of particles in these theories and then 
follow a TBA procedure
to derive a set of thermodynamics equations, eq.~(\ref{order-k-eq}), 
which describes what we call `order-$k$ non-abelian 
exclusion statistics'. We shall show that these equations can be
used to rederive the known thermodynamic equations for a variety of 
non-abelions. 

The approach followed in this paper can be summarized as follows. The
essence of non-abelian statistics is a degeneracy of the quantum state for
a number of quasi-particles with all labels (quantum numbers and
positions) fixed. In the setting of a CFT approach, this degeneracy can be
linked to a choice of fusion path in a multi-quasi-particle state
\cite{fendley,BLS2,MR,NW}. In the context of an associated scattering
problem, non-trivial fusion rules lead to non-diagonal scattering, and
under a TBA procedure, the latter leads to an extended TBA system that
features a number of non-physical  pseudo-particles. In this paper we
give a general characterization of the extended TBA system for
quasi-particles satisfying order-$k$ non-abelian exclusion statistics. 
Specializing to concrete examples of interest, we demonstrate that this 
extended TBA
system reduces to equations that were previously obtained by other
methods.  Our main example will be that of spinons for spin $S\geq 1$
integrable spin chains, which correspond to the CFT associated to
$SU(2)_{k=2S}$. 

As an application of the formalism presented here, we shall present the
defining equations for the thermodynamic distribution functions for
quasi-hole excitations over the non-abelian quantum Hall states recently
proposed by N.~Read and E.~Rezayi \cite{RR}.

The structure of the extended TBA system for non-abelions is reflected in
the structure of so-called fermionic sum formulas for the characters of an
associated CFT. We shall explain this connection, again using $SU(2)_k$
spinons as our guiding example.

In the study of CFT quasi-particles, one encounters non-abelian exclusion
statistics that are more general than the `order-$k$ non-abelian exclusion
statistics' that we study in this paper. [An explicit example is provided
by the quasi-holes over the non-abelian spin-singlet quantum Hall states 
recently proposed in \cite{AS}.] The approach followed in this paper can 
be generalized to such more complicated situations \cite{BC}.

\section{Abelian statistics: thermodynamic equations and TBA}

In his definition of `fractional exclusion statistics'
Haldane \cite{Ha2} makes the assumption that {\it the act of 
filling a single particle state (by a particle of type $a$)
reduces the dimensionality of the space of states available
to a particle of type $b$ by the amount $G_{ab}$}, with 
${\bf G}$ a so-called statistics matrix.
At the level of thermodynamics, Haldane's assumption leads to 
a set of equations for the single-level grand canonical
partition functions $\la_a = \la_a(z_1,\ldots,z_n)$, with $z_a=
e^{\beta(\mu_a-\epsilon^{(0)})}$, with $\epsilon^{(0)}$
the (bare) energy and $\mu_a$ the chemical potential for
particles of type $a$ \cite{WuIs, FuKa}, 
\be
\left( { \la_a-1\over \la_a}\right) \prod_b \la_b^{G_{ab}} = z_a \,,
\label{wu-eq}
\ee
from which the 1-particle distribution functions can obtained as 
\be
n_a(\epsilon^{(0)}) =
z_a {\partial \over \partial z_a} \log \lambda(z)  _{\big|
z_b = e^{\beta(\mu_b-\eps^{(0)})}}  \,.
\label{dist}
\ee
In view of the relation, in the context of quantum Hall systems, 
between the exclusion statistics leading to eq.~(2.1) and 
abelian braid statistics, we shall call the statistics underlying
eq.~(\ref{wu-eq}) `abelian exclusion statistics'.

In the context of the Thermodynamic Bethe Ansatz (TBA), statistics
properties are derived from scattering data. 
For a factorizable $S$-matrix with diagonal 2-particle scattering 
matrix $S_{ab}(\theta)$, the thermodynamics follow from the following
TBA equations for dressed energies $\epsilon_a(\theta)$
\be
\eps_a(\theta) = (\eps^{(0)}(\theta) - \mu_a) - {1 \over \beta} \sum_b
(\phi_{ab} * \ln(1+e^{-\beta \eps_b}))(\theta)
\label{TBA-eq}
\ee
where * denotes the convolution and  $\phi_{ab}(\theta) = -i
\del_{\theta} \ln S_{ab}(\theta)$.
One easily checks \cite{BW} that for an 
$S$-matrix of the form 
\be
S_{ab}(\theta) = \exp[ i (\delta_{ab}-G_{ab}) \Theta(\theta)] \ .
\label{SU(2)_1}
\ee
the TBA equations (\ref{TBA-eq}) are equivalent to the 
equations  (\ref{wu-eq})  for 
the grand partition functions $\lambda_a=1+e^{-\beta\eps_a}$.

It has been found that in some of the simplest integrable many-body systems 
with inverse square interactions, such as the spin-${1 \over 2}$ 
Haldane-Shastry chain, the $S$-matrix for fundamental 
excitations precisely takes the `statistical' form  of 
eq.~(\ref{SU(2)_1}) \cite{Ha1}. Such systems allow an interpretation 
as `free gases of fractional statistics particles'. 
[See \cite{PT} for another example of a system with purely 
statistical interactions]. In more general integrable systems, 
such as the spin $S={1 \over 2}$ Heisenberg chain, one finds 
$S$-matrices that agree with (\ref{SU(2)_1}) only in the limit of 
large $\theta$. We refer to \cite{Es} for a discussion.

\section{Spinons in the $SU(2)_k$ WZW theory}

It has been observed \cite{Ha1,level-1,BLS1}, that the integrable 
structure of the spin-${1 \over 2}$ Haldane-Shastry chain can be 
carried over to the associated $SU(2)_1$ CFT. As a consequence, 
one may study the latter CFT in terms of spin-$\half$ quasi-particles 
$\phi_{\uparrow}$, $\phi_{\downarrow}$ with scattering matrix of 
the type (\ref{SU(2)_1}), with $G_{ab}=\half$ for $a,b=\uparrow,\downarrow$.

For a similar analysis of $k>1$ $SU(2)_k$ WZW models, one would like to
start from a $S>\half$ version of the Haldane-Shastry spin chain, but such
theories have not been constructed. What one may do instead \cite{FS}, is
study the integrable Heisenberg chains with $S>\half$, and analyze the way 
the spin-$\half$ spinons build the excitation spectrum. Proceeding in this 
manner, the authors of \cite{FS} have made contact with
an approach which starts from truncations of the conformal spectrum and
used the recursion method of \cite{Sc1} to derive the thermodynamic 
equations for the spinon excitations.

In this paper we offer a third point of view on the statistics of
$SU(2)_k$ spinons. We shall start from scattering data for these
quasi-particles and employ the Thermodynamic Bethe Ansatz to derive the
thermodynamics.  We shall find agreement with the results of \cite{FS},
and we shall be able to generalize the result obtained to a general set of
thermodynamic equations for quasi-particles satisfying what we call
`order-$k$ non-abelian exclusion statistics'.

The TBA equations derived below are closely related to a special limit of
TBA equations that have been studied in
\cite{reshetikhin,zamolodchikov,fendley}. In our presentation we shall
start by closely following these references, and then proceed to make
the connection with non-abelian exclusion statistics.
 
The $SU(2)_k$ spinons are massless particles which transform as doublets
($\uparrow,\downarrow$) under $SU(2)$. In addition, these particles have a
kink structure with vacua labeled by an index running from $1$ to $k+1$.
In the CFT setting, the origin of these vacua is the following.  The
chiral CFT spectrum decomposes into $k+1$ sectors, which are each headed by
a primary field of the affine $SU(2)_k$ symmetry. The primary fields are
labeled by their $SU(2)$ spin $j$ for which the allowed values are $j=0,
\half,1,\ldots,{k \over 2}$. The spinon is associated to the spin-$\half$
Chiral Vertex Operator (CVO), which interpolates between adjacent vacua. 

In a scattering picture, the kinks scatter off adjacent kinks in a
non-diagonal manner. The $S$-matrix will be written as $S = S_{\uparrow,
\downarrow} \otimes S_{\rm kink}$, with the first factor identical to the
level-1 statistical scattering matrix (i.e., eq.~(\ref{SU(2)_1}) with
$G_{ab}=\half$). For the kink part of the $S$-matrix  we propose the 
`statistical limit' of the ${(\rm RSOS)}_{k+1}$ solution of the Yang-Baxter
equations as described in \cite{zamolodchikov}. Following this reference,
we shall label the kinks as $B_{\alpha \beta}(\theta)$, where $\theta$ is
the kink rapidity and $\alpha, \beta$ are the labels of two adjacent vacua. 
The scattering of adjacent kinks is represented as 
\be B_{\alpha
\gamma}(\theta_1) B_{\gamma \beta}(\theta_2) = \sum_\delta S^{\gamma
\delta}_{\alpha \beta}(\theta_1-\theta_2) B_{\alpha \delta}(\theta_2) 
B_{\delta \beta}(\theta_1)   
\label{def-s-matrix}  \ .
\ee
In the subsections that follow, we shall present an explicit $S$-matrix 
for spinons in the $SU(2)_2$ theory, and discuss the TBA equations for
spinons in the $SU(2)_k$ theory.

\subsection{Scattering matrix for spinons in the $SU(2)_2$ theory}

For $({\rm RSOS})_3$, there are four kinds of kinks, $B_{0
, \pm}$ and $B_{\pm,0}$ and the non-zero scattering amplitudes are
$S_{00}^{\alpha \beta}$ and $S_{\alpha \beta}^{00}$, with $\alpha,\beta=+,-$. 
The $S$-matrix is a $ 6 \times 6$ matrix, and it satisfies the conditions 
for crossing symmetry (\ref{crossing-symmetry}),
unitarity (\ref{unitarity}), and factorization (\ref{factorization})
\be
S_{00}^{\alpha \beta}(\theta) = S_{\alpha \beta}^{00}(i \pi - \theta)
\label{crossing-symmetry}
\ee
\be
\sum_{\gamma=\pm} S_{00}^{\alpha \gamma}(\theta)S_{00}^{\gamma
\beta}(- \theta) = \delta_{\alpha \beta} \ ,
\qquad
S_{\alpha \beta}^{00}(\theta)S_{\alpha \beta}^{00}(- \theta) =1 
\label{unitarity}
\ee
\be
\sum_{\gamma} S_{00}^{\alpha \gamma}(\theta)S_{\beta \gamma}^{00}(
\theta + \theta')S_{00}^{\delta \gamma}(\theta')=S_{\beta \delta}^{00}
(\theta)S_{00}^{\alpha \delta}(\theta + \theta')S_{\alpha
\beta}^{00}(
\theta')
\label{factorization}
\ee

The $S$-matrix that is consistent with the above conditions is of the form
\be
S^{\alpha \beta}_{00}(\theta) = e^{-i \rho \theta} f_{\alpha \beta}
\, \sigma(\theta)
\ ,
\qquad
S^{00}_{\alpha \beta} (\theta) = e^{i \rho \theta} g_{\alpha \beta} 
\, \sigma(\theta)
\ee
where
\be 
\sigma (\theta) = {1 \over [{\cosh(\theta/2)}]^{1/2}} \exp [{i \over 4}
I(\theta)] \ , 
\qquad  
I(\theta)= {\rm sign} (\theta) \int_0^\infty {dt \over t} 
{ \sin {|\theta| t \over \pi} \over \cosh^2{t \over 2} }
\label{s-matrix}
\ee
and $\rho= {1 \over 2 \pi} \log2$.

To obtain the kink part $S_{\rm kink}$ of the  $S$-matrix
of $SU(2)_2$ spinons, we take the `statistical limit'
$\theta \rightarrow \pm \infty$ of the $({\rm RSOS})_3$ $S$-matrix. 
In this limit we have
\be
I(\theta) \rightarrow {\rm sign}(\theta) 
\int_0^\infty  dt {\sin t \over t} 
= {\pi \over 2}
{\rm sign}(\theta) 
\label{limit-integral}
\ee
so that
\be
\sigma(\theta) \rightarrow \sqrt{2} \, e^{\mp {\theta \over 4}} 
e^{\pm {\pi i\over 8}}
\label{limit-sigma}
\ee
The matrix elements $ f(\theta)$ and $g(\theta)$ given in 
\cite{zamolodchikov} have the limiting form  
\bea
f_{++} (\theta)= f_{--}(\theta)
\rightarrow 
{1 \over 2} e^{\pm {\theta \over 4}}
\ , &\quad&
f_{+-}(\theta)= f_{-+}(\theta) 
\rightarrow  
\pm {i \over 2} e^{\pm {\theta \over 4}}
\nonu
g_{++}(\theta) = g_{--}(\theta) 
\rightarrow {1 \over 2} e^{\pm {\theta \over 4}} (1 \mp i)
\ , &\quad&
g_{+-}(\theta) = g_{-+}(\theta)
 \rightarrow
{1 \over 2} e^{\pm {\theta \over 4}} (1 \pm i) \ .
\eea
The  $S$-matrix  elements are the following when $\theta \rightarrow \pm
\infty$
\bea
 S_{00}^{++} = S_{00}^{--}= { a^{-1}(\theta) \over \sqrt 2}
e^{ \pm {\pi i\over 8}}
\ , &\quad&
S_{00}^{+-} = S_{00}^{-+} = \pm i { a^{-1}(\theta) \over \sqrt 2}
e^{ \pm {\pi i\over 8}}
\nonu
S_{-+}^{00}=S_{+-}^{00}= a(\theta) e^{\pm {3 \pi i \over 8} }
\ , &\quad&
S_{++}^{00}=S_{--}^{00}= a(\theta)  e^{\mp { \pi i \over 8}}
\label{s-matrix3}
\eea
where $a(\theta)=  ({\sqrt 2})^{{i \theta \over \pi}} $ is a gauge factor.

\subsection{TBA for spinons in the $SU(2)_k$ theory}

To write down the TBA equations for the $SU(2)_k$ spinons, we must 
consider an ensemble of $N$ spinons and allow one of them to scatter 
with all the
others, taking the thermodynamic limit $N \rightarrow \infty$ in the end. 
The scattering in the $(\uparrow \downarrow)$ part is diagonal
whereas 
in the RSOS part the 
scattering is non-diagonal and therefore we have to diagonalize it before
quantizing the system. Since the $S$-matrix of $SU(2)_k$ is a tensor
product of the $(\uparrow \downarrow)$ and the RSOS part, we first
diagonalize the RSOS part before putting it together with the 
$(\uparrow \downarrow)$ part.  

The RSOS scattering can be described by a transfer matrix 
$T(\theta)$ which has the entries 
\be
T_{\alpha_1',
\alpha_2'...\alpha_N'}^{\alpha_1, \alpha_2,...\alpha_N} (\theta| \theta_1,
\theta_2,...\theta_N)
 = S_{\alpha_1 \alpha_2'}^{\alpha_1' \alpha_2}(\theta
-\theta_1) S_{\alpha_2 \alpha_3'}^{\alpha_2' \alpha_3}(\theta
-\theta_2)
\ldots S_{\alpha_N \alpha_1'}^{\alpha_N' \alpha_1}(\theta
-\theta_N) \ .
\label{factorized}
\ee
The diagonalization of $T(\theta)$ involves finding the
eigenvalues $\Lambda_i(\theta|\theta_1,\theta_2,...\theta_N)$ and
eigenvectors $\psi_i$:
\be
T(\theta|\theta_1,\theta_2,...\theta_N)
\psi_i=\Lambda_i(\theta|\theta_1,\theta_2,...\theta_N) \psi_i  \ .
\ee
Quantizing the particles in a finite system of size $L$
requires 
the eigenvalues $\Lambda$ to satisfy the condition
\be
e^{i p_l L} \Lambda(\theta_l| \theta_1,...\theta_N)=1 \ ,
\qquad l=1,2,...N.
\ee

The TBA equations for $({\rm RSOS})_{k+1}$ have been given in 
\cite{zamolodchikov} with a detailed derivation of the diagonalization 
of the $k=2$ case. We will not repeat the details here. The final TBA system
describes a single physical particle, labeled as $i=0$, and $k-1$ auxiliary
`pseudo-particles', labeled as $i=1,2,\ldots,k-1$, which arise from the 
diagonalization of the transfer matrix. Denoting the densities of the 
occupied kink states by $\rho_i(\theta)$ and the density of available 
states as $P_i(\theta)$, we can parametrize
\be
{\rho_i \over P_i}= {e^{-\epsilon_i} \over 1+ e^{-\epsilon_i}} \ .
\ee  
The dressed energies $\epsilon_i$, $i=0,1,...k-1$,  satisfy an effective 
TBA system \cite{zamolodchikov} of the form eq.~(\ref{TBA-eq}), in which 
the particles $i=0,1,...,k-1$ scatter off each other purely diagonally with 
$S$-matrices described by TBA kernel 
\be
\phi_{ij}(\theta)= {1 \over 2\pi} \ {1\over \cosh\theta } \ l_{ij} \ ,
\label{TBA-kernel}
\ee
where $l_{ij}$ is the adjacency matrix of the $A_{k}$ dynkin
diagram. In this TBA system, the physical particle has a standard
bare energy term (the term $\epsilon^{(0)}-\mu_0$ in
eq.~(\ref{TBA-eq})), 
but there is no such term for the pseudo-particles $i=1,\ldots,k-1$.
We remark that in an approach based on a Bethe Ansatz solution of a
lattice (RSOS)$_{k+1}$ model, the pseudo-particles correspond to
the (complex) string solutions to the Bethe equations.

In the statistical limit ($\theta \to \pm \infty$) appropriate for
$SU(2)_k$ spinons, the kernel eq.~(\ref{TBA-kernel}) tends to
\be
\phi_{ij}(\theta) \to \half \,  \delta(\theta) \, l_{ij} .
\ee

The TBA system for the $(\uparrow \downarrow)$ part of the $S$-matrix
is that of the
$SU(2)_1$ spinons. It can be written as
\be
\left( {\lambda_\uparrow - 1 \over \lambda_\uparrow}\right) 
\lambda_\uparrow^{\half} \lambda_\downarrow^{\half}
= z_\uparrow \ ,
\qquad
\left( {\lambda_\downarrow - 1 \over \lambda_\downarrow}\right) 
\lambda_\uparrow^{\half} \lambda_\downarrow^{\half}
= z_\downarrow \ ,
\label{su2-level1-TBA}
\ee
with $\lambda_A=1+e^{-\epsilon_A}$ and $z_{A}
=e^{\beta(\mu_A- \eps^{(0)})}$ for $A=\uparrow,\downarrow$.

The TBA systems for the $(\uparrow \downarrow)$ part and the RSOS part,
which have been solved individually, should now be combined.
The transfer matrix for the combined system is built from
the tensor product of the $(\uparrow \downarrow)$ and the RSOS parts.  
The eigenvalues of the transfer matrix for the $SU(2)_k$ system,  
$\Lambda_i(\theta| \theta_i, \theta_2,....\theta_n)$, are products of the 
eigenvalues of the $(\uparrow \downarrow)$ and the $({\rm RSOS})_{k+1}$ 
part.  The TBA kernel for the combined system is 
the sum of those of the $(\uparrow \downarrow)$ and the RSOS ones.

In the combined TBA system, the place of the $i=0$ physical particle
in the (RSOS)$_{k+1}$ system is taken by the physical particles
which we denoted as $\phi_{\uparrow,\downarrow}$. The total particle
content is thus labeled as $a=\uparrow,\downarrow,1,\ldots,k-1$.
The combined TBA equations take the following form
\bea
\left( {\lambda_\uparrow - 1 \over \lambda_\uparrow}\right) 
\lambda_\uparrow^{\half} \lambda_\downarrow^{\half} 
\lambda_1^{-\half}
= z_\uparrow  \ , &\quad&
\left( {\lambda_\downarrow - 1 \over \lambda_\downarrow}\right) 
\lambda_\uparrow^{\half} \lambda_\downarrow^{\half} 
\lambda_1^{-\half}
= z_\downarrow \ ,
\nonu
\left( {\lambda_1 - 1 \over \lambda_1}\right) 
\lambda_\uparrow^{-\half} \lambda_\downarrow^{-\half} 
\lambda_1 \lambda_2^{-\half}
=1 \ , &\quad&
\left( {\lambda_j - 1 \over \lambda_j}\right) 
\lambda_{j-1}^{-\half} \lambda_j \lambda_{j+1}^{-\half} = 1 
\ , \quad  j=2,\ldots,k-1 \ ,
\nonu
\label{su2-k-TBA}
\eea
with $\lambda_a = (1 + e^{-\epsilon_a(\theta)})$ for $a=\uparrow, 
\downarrow,1,\ldots,k-1$.

As a check on the correctness of our assignment of $S_{\rm kink}$, we
have compared the TBA system (\ref{su2-k-TBA}) with results obtained from 
different approaches. Starting with $k=2$, we have
\bea
&& \left( {\lambda_\uparrow - 1 \over \lambda_\uparrow}\right) 
\lambda_\uparrow^{\half} \lambda_\downarrow^{\half} 
\lambda_1^{-\half}
=z_\uparrow
\ , \quad
\left( {\lambda_\downarrow - 1 \over \lambda_\downarrow}\right) 
\lambda_\uparrow^{\half} \lambda_\downarrow^{\half} 
\lambda_1^{-\half}
= z_\downarrow
\nonu
&& \left( {\lambda_1 - 1 \over \lambda_1}\right) 
\lambda_\uparrow^{-\half} \lambda_\downarrow^{-\half} \lambda_1
= 1 \ .
\label{su2-level2-TBA}
\eea
Eliminating $\lambda_1$, putting $x=z_{\uparrow}=
z_{\downarrow}$, and defining 
$\lambda=\lambda_{\uparrow} \lambda_{\downarrow}$, one finds
\be
(\lambda^{\half}
-1)^2-x^2(\lambda^{\half}
+1)=0 .
\ee
This equation is in agreement with the 
results of \cite{FS,BS2}, which were obtained through an entirely 
different 
approach. The corresponding equations for $k=3,4$, which were 
explicitly given in \cite{BS2}, are recovered in a similar manner.

It was argued in \cite{Sc2,BS2} that, in general, the non-abelian nature 
of the exclusion statistics of a set of physical particles manifests 
itself in the dependence of the grand partition function $\lambda(z_a)$ 
on the variables $z_a$, in the limit where $z_a \rightarrow 0$ 
(corresponding to the Boltzmann tails of the associated thermodynamic 
distribution functions). 
Expanding the partition function as
\be
\lambda(z) = 1 + \alpha_a z_a + \ldots
\ee
one observes \cite{Sc2,BS2} that the degeneracies that are  characteristic 
for non-abelian statistics lead to coefficients $\alpha_a >1$. It is easily 
seen that the equations (\ref{wu-eq}) for abelian exclusion statistics 
lead to $\alpha_a=1$, 
for all choices of the statistics matrix ${\bf G}$. In contrast, the 
TBA equations (\ref{su2-k-TBA}) lead to 
\be
\alpha_\uparrow= \alpha_\downarrow = 2 \cos({\pi \over k+2}) \ .
\label{entr-fac}
\ee
In earlier work \cite{fendley,BS2}, the factors (\ref{entr-fac}) were 
obtained as the largest eigenvalues of the fusion 
matrix associated with a single spinon field. In the literature on rational 
CFT, these factors are known as `quantum dimensions' associated to 
specific conformal fields.

\section{TBA for order-$k$ non-abelian exclusion statistics}

Having understood the example of $SU(2)_k$ spinons, we can generalize the 
TBA system (\ref{su2-level2-TBA}) to more general particles that satisfy 
an analogous form of exclusion statistics, which we call
`order-$k$ non-abelian exclusion statistics'. We propose the following 
equations for a set of $n$ physical particles, labeled as $A=1,2,\ldots,n$
together with $k-1$ pseudo-particles labeled as $i=1,\ldots,k-1$ 
\bea 
\left( { \la_A-1\over \la_A}\right) 
\prod_B \la_B^{\widetilde{G}_{AB}} 
\prod_i \la_i^{\widetilde{G}_{Ai}} = z_A && \,, \quad A=1 \ldots,n
\nonu 
\left( { \la_i-1\over \la_i}\right) 
\prod_A \la_A^{\widetilde{G}_{iA}} 
\prod_j \la_i^{\widetilde{G}_{ij}} = 1 && \,, \quad  i=1,\ldots,k-1 \ ,
\label{order-k-eq}
\eea
with the matrix $\widetilde{G}_{AB}$ specifying the
`abelian part' of the statistics, 
$\widetilde{G}_{Ai}=\widetilde{G}_{iA}= -\half \delta_{i,1}$
and ${\widetilde{G}}_{ij}=\half (C_{k-1})_{ij}$, with $C_{k-1}$
the Cartan matrix of $A_{k-1}$. In specific examples, one may
reduce (\ref{order-k-eq}) to an equation for $\lambda=\Pi_A\lambda_A$, 
which can then be compared  with results obtained by other methods.

Comparing with section 2, we see that the extended TBA system
(\ref{order-k-eq}) describes a situation where Haldane's exclusion
principle is applied to a system of both physical and auxiliary
(pseudo-) particles. Despite the similarity between the `abelian'
equations (\ref{wu-eq}) and the `non-abelian' extension
(\ref{su2-k-TBA}), there is an essential difference in the physics 
that is described. In physical terms, one could say that the absence 
of the bare energy term in the TBA equations for the pseudo-particles 
($i=1,\ldots,k-1$) means that these particle are not 
suppressed at high energies and betray their presence by combinatorial 
factors that are characteristic of non-abelian statistics.

Clearly, the TBA system (\ref{order-k-eq}) has been obtained from that 
for $SU(2)_k$ spinons by changing the abelian part of the scattering but 
keeping the `kink-structure'. In CFT terms this means that in all cases 
the 
general TBA system (\ref{order-k-eq}) will refer to conformal fields with 
fusion rules identical to those of $SU(2)_k$ spinons. These fusion rules 
are, for example, found in CFTs describing non-abelian quantum Hall 
states and 
in minimal models of the Virasoro algebra, and we shall see that these 
provide other physically meaningful examples of the general structure 
(\ref{order-k-eq}).

The robustness of the `kink-structure' behind the TBA system 
(\ref{order-k-eq}) implies that the combinatorial implications of the 
presence of the pseudo-particles are shared by all realizations. This 
holds in particular for the `entropy' factors (\ref{entr-fac}), which 
are universal.
The importance of the entropy factors (\ref{entr-fac}) has been stressed 
in other contexts, in particular in that of the $k$-channel Kondo model
\cite{fendley}.

\section{Non-abelian quantum Hall states}

In a recent paper \cite{Sc2}, one of us studied the exclusion statistics 
properties of quasi-hole excitations over the so-called $q$-pfaffian 
quantum Hall states.
The CFTs that describe the edge excitations over these quantum Hall states 
contain, in addition to the free boson associated to edge magnetoplasmons,
a free Majorana fermion that is associated to an additional  `dipole' degree 
of freedom. The conformal fields that create the quasi-hole excitations 
contain as a factor the spin-field associated to the Majorana fermion. The 
fusion rules of this spin field are identical to those of the $SU(2)_2$ 
spinon fields and, by the general reasoning that we presented, we expect 
that the quasi-holes will provide a realization of order-2 non-abelian 
statistics.
In this situation, the abelian part $\widetilde{G}_{AB}$ of the statistics 
matrix is simply a number, which is set by the value of $q$. We find
the TBA system
\be
\left( \lambda-1 \over \lambda \right) \lambda^{{q+1 \over 4q}} 
\lambda_1^{-1/2}  = z
\ ,  \qquad
\left( \lambda_1-1 \over \lambda_1 \right) \lambda^{-1/2} 
\lambda_1=1
\ee
which, upon eliminating $\lambda_1$, is indeed equivalent to the result of 
\cite{Sc2}.

In a recent paper \cite{RR}, the $q$-pfaffian quantum Hall  
states have been generalized to a class of non-abelian quantum Hall states 
that are parametrized by integers $(k,M)$. These states have filling 
fraction $\nu=k/(Mk+2)$. The guiding principle in the construction of 
these states 
has been the replacement of the Majorana fermion in the pfaffian CFT by 
an order-$k$ parafermion in the sense of CFT. The fundamental quasi-holes 
over the $(k,M)$ states are associated to conformal fields that contain 
as a factor the spin-field associated to an order-$k$ parafermion, and 
it is a well-known fact that these spin fields have the same fusion rules 
as the spinons in an $SU(2)_k$ WZW theory. We therefore expect that the 
exclusion statistics of the quasi-holes over the Read-Rezayi states are 
of the general form (\ref{order-k-eq}), with a single abelian component
(see also \cite{FNS}). 
The only number that needs to be determined is the exponent $g(k,M)$ in 
the leading equation
\be
\left( \lambda-1 \over \lambda \right) \lambda^{g(k,M)} \lambda_1^{-1/2}  
= z \ .
\ee
By matching the thermodynamic content of the TBA system with known 
information (in particular, the value of the Hall conductance), 
we determined 
\be
g(k,M) = {(k-1)M+2 \over 2(kM+2)} \ .
\ee
The $q$-pfaffian arises as the special case $k=2$, $M=q-1$ of the more 
general construction, and this provides a check on the result for $g(k,M)$.

We remark that the (positively charged) quasi-holes by themselves do not 
generate the full (chiral) spectrum of the edge CFT for the Read-Rezayi 
states. As explained in \cite{ES,Sc2}, one may identify (negatively
charged) quanta that are dual to the quasi-holes, and that satisfy a
form of exclusion statistics that is dual to the statistics of the
quasi-holes. Upon combining the two kinds of excitations, one recovers
the full chiral spectrum. For the (abelian) Laughlin states ($k=1$)
and for the $q$-pfaffian states ($k=2$) such a result has been 
explicitly demonstrated in \cite{ES,Sc2}.

\section{Generalized fermions in CFT}

A third example of conformal fields with the fusion rules
of $SU(2)_k$ spinons are the fields labeled as $\Phi_{(2,1)}$
in the unitary minimal model ${\cal M}^{k+2}$ of the Virasoro 
algebra. These CFTs have central charge $c(k)=1-{6 \over 
(k+2)(k+3)}$ and the field $\Phi_{(2,1)}$ has scaling dimension
$\Delta_{2,1}={k+5 \over 4(k+2)}$. For $k=1$ the field
$\Phi_{(2,1)}$ is the Majorana fermion of the Ising CFT, while
in the limit $k\to\infty$, $\Phi_{(2,1)}$ is identified 
with one of the spinons of the $SU(2)_1$ WZW theory.

In \cite{BS2}, the grand partition function $\lambda(x)$ 
was studied on the basis of recursion relations satisfied by 
truncated conformal characters, and explicit polynomial
equations for $\lambda(x)$ were obtained for low $k$.
We remark that recursion relations similar to those 
employed in \cite{BS2}, have been used in the analysis
by Andrews, Baxter and Forrester (ABF) of local height 
probabilities in specific RSOS models \cite{ABF} 
(see also \cite{ferchar2}). Interestingly, the two situations
are `dual' in the sense that where the ABF recursions are in
terms of the system size $m$, the recursion relations
for the `truncated spectra' of \cite{BS2} are in terms
of a momentum (or energy) variable $l$. 

Following the general arguments presented in section 4 of
this paper, we expect that the exclusion statistics of the 
$\Phi_{(2,1)}$
quanta is described by an order-$k$ extended TBA system
of the form (\ref{order-k-eq}). The abelian part of the
TBA system has a single component ($A=1$), and we checked
that for $k=1,2,3$, with the assignment $\widetilde{G}_{11}=1$,
the extended TBA system (\ref{order-k-eq}) reduces to
the result obtained in \cite{BS2}. For $k=1$ the value
$\widetilde{G}_{11}=1$ simply reflects the fermionic statistics
of the Majorana fermion.

We remark that the full matrix ${\bf \widetilde{G}}$ matrix is 
equal to one half times the Cartan matrix $C_k$ of $A_k$ and that 
the extended TBA system for the $\Phi_{(2,1)}$ quanta in the 
minimal model ${\cal M}^{k+2}$ is identical to the TBA system 
for (RSOS)$_{k+1}$ in its statistical limit (see section 3.2). 
This observation nicely fits with the fact that the scattering 
matrix for the $\Phi_{(1,3)}$ massive perturbations of ${\cal M}^{k+2}$ 
exhibits a similar (RSOS)$_{k+1}$ structure \cite{Za}. 
Taking the limit $k\to\infty$, one finds that $\half C_{\infty}$
acts as the statistics matrix in the $SU(2)_1$ CFT. This result 
was recently discussed in the context of the Bethe Ansatz solution
of the (single channel) Kondo model \cite{PV}.

The form of the extended TBA system for the generalized 
fermions is further confirmed by an inspection of the `fermionic
sum formulas' for the conformal characters. This connection
is the subject of our next section.

\section{Relation with `fermionic' character formulas}

A beautiful development in the mathematical analysis of CFT
has been the work, initiated in \cite{3-potts}, on so-called 
fermionic 
sum formulas for the characters in a variety of minimal models
of CFT. In the initial work, these expressions had their origin
in the mathematical structure of the space of solutions to the 
Bethe equations for a specific integrable 3-state Potts chain. 
The character identities that resulted from this analysis have 
been generalized to large classes of models of CFT
\cite{ferchar1}, and many of the results have been put on a rigorous 
mathematical basis \cite{ferchar2}. On the basis of this extensive
work, a general fermionic sum formula of the form
\be
\sum_{ \stackrel{\scriptsize\bf m=0}{\rm restrictions} }^{\infty}
q^{ {1 \over 2}{\bf mBm} - {1 \over 2}{\bf Am} } \,
\prod_{\alpha=1}^n y_{\alpha}^{m_\alpha} \,
\prod_{\alpha=1}^n 
\left[ \begin{array}{c}
( (1-{\bf B}){\bf m} + { {\bf u} \over 2} )_{\alpha} \\
m_\alpha \end{array} \right]
\label{fer}
\ee
has been put forward \cite{BM}. Important ingredients in this 
expression are the summation variables $m_1,\ldots,m_n$, a bilinear 
form set by an $n\times n$ matrix ${\bf B}$ and a $n$-vector of 
parameters ${\bf u}$. The square brackets $[\ldots]$ denote
the $q$-binomial coefficient.

In an independent development, it has been recognized that in 
many CFTs it is possible to construct a quasi-particle basis 
directly in terms the conformal fields, i.e. without making 
reference to an underlying integrable model. The proto-typical
example for this is the spinon basis for the $SU(2)_1$ CFT,
which has been fully justified on the basis of
generalized commutation relations satisfied by modes of the 
conformal spinon fields \cite{BLS1}. In the paper \cite{BLS1}, 
the $SU(2)_1$ spinon basis results were explicitly linked with some 
of the character identities obtained in \cite{ferchar1}.
In further studies along these same lines, spinon-type quasi-particle 
bases for higher level ($SU(2)_k$) \cite{BLS2,NY} and higher rank 
($SU(n)_1$) \cite{BS0,KKN} WZW models have been constructed, and 
associated character identities were obtained. 

In a 1997 Letter \cite{Sc1}, one of us has proposed that the 
quasi-particle bases that are constructed in terms of specific 
conformal fields can be used to assign a form of exclusion statistics 
(in the sense of Haldane and more general) to the quanta of these
same conformal fields. Again, the $SU(2)_1$ spinons served as a 
prototypical 
example. In the paper \cite{BS2}, a systematic study has been 
presented and many new forms of exclusion statistics have been 
revealed. Of particular interest have been the applications to edge 
theories for various quantum Hall systems \cite{ES,Sc2,FNS}, where 
other, closely related, notions of fractional statistics have been 
studied in detail.

In particular cases, there is a close connection between the 
fermionic sum type character expressions (\ref{fer}) and 
the exclusion statistics associated to conformal fields.
In particular, it has been observed \cite{Hik,ES} that
for a single particle species satisfying Haldane's statistics
with parameter $g$, the partition sum is naturally of the form
(\ref{fer}) with $B_{11}=g$ and $u_1=\infty$. This connection
was put on a general footing in a conjecture put forward
in \cite{Sc3,BM}. Focusing on particles satisfying abelian
statistics with statistics matrix ${\bf G}$, one expects a 
direct connection between the TBA system (\ref{wu-eq}) and
the fermionic sum (\ref{fer}) with ${\bf B}={\bf G}$ and
all $u_a=\infty$. See, eg, \cite{BS2}, where the associated
central charge identity was explicitly checked.

The result obtained in this paper allow us to illustrate the
connection between fractional statistics and fermionic sum 
formulas in the non-abelian case.
For some of the CFT non-abelions that we treated in the above,
explicit fermionic sum formulas are known in the literature,
and one quickly recognizes how the data in the extended TBA 
systems of the form (\ref{order-k-eq}) can be `matched' with 
the data contained in the fermionic sum formulas. In particular, 
one finds that the full statistics matrix ${\bf \widetilde{G}}$ 
is to be identified with the bilinear form ${\bf B}$ in the 
leading $q$ power, while the values of $u_a$ distinguish between 
physical particles ($u_A=\infty$) and pseudo-particles 
($u_i<\infty$).

For an explicit example, we recall the spinon form of the 
characters of $SU(2)_k$ affine Kac-Moody algebra in the 
corresponding WZW model \cite{BLS2,NY}. The character of the
highest weight module built over the primary state of spin 
$j \leq k/2$ is written as
\be
{\rm ch}_j(z,q)
=
q^{\Delta_{(j)} - j/2}
\ 
\sum_{M,N\geq 0}
\ 
q^{-{1\over 4} (M+N)^2}
\ 
\Phi^{M+N}_{C_k}(u_j, q)
\ 
{z^{\half(M-N)} \over (q)_M (q)_N}
\ee
with 
\be
\Phi^{m_1}_{K}({\vec u}, q)
= \sum_{m_2, ..., m_k} \ 
q^{ {1\over 4} {\vec m} \cdot K \cdot {\vec m} \  }
\prod_{i\geq 2}
\biggl [
\begin{array}{c}
{1\over 2}
{ \bigl (
(2-K ) \cdot {\vec m} + {\vec u}
\bigr)}_i
\\
m_i 
\end{array}
\biggr ]
\ee
with the vector ${\bf u_j}$ given by $(u_j)_i=\delta_{i,2j+1}$
and $C_k$ equal to the Cartan matrix of $A_k$ (see \cite{BLS2}
for further details).
The structure of this formula is closely related to the 
factorized structure of the extended TBA system. The numbers 
$M,N$ stand for the number of $\phi_\uparrow,\phi_\downarrow$ 
quanta, and the numbers $m_1,\ldots,m_k$ correspond to
the particle content of the (RSOS)$_{k+1}$ kink sector.
The non-trivial `gluing' of the two sectors is implemented
by the identification $m_1=M+N$. One quickly checks that
the resulting bilinear form (on the $2+(k-1)$ variables 
$M,N,m_2,\ldots,m_k$) agrees with the matrix 
${\bf \widetilde{G}}$ in the TBA system (\ref{su2-k-TBA}). 

The extended TBA system for the generalized fermions of 
section 6 can, in a very similar way, be recognized in the 
structure of particular fermionic sum formulas for the 
conformal characters of the ${\cal M}^{k+2}$ minimal
model. These formulas were first proposed in 
\cite{ferchar1}, and they have been proven in \cite{ferchar2}. 
The appropriate bilinear
form, for summation variables $m_1,\ldots,m_k$, agrees with
the matrix ${\bf \widetilde{G}}$ identified in section 6 and
via the values of the $u_a$ the particle labeled as $a=1$
is singled out as the only physical particle. 

\section{Conclusion}

In this paper, we have proposed a general extended TBA system,
eq.~(\ref{order-k-eq}), 
for physical particles satisfying what we call order-$k$ 
non-abelian exclusion statistics. This TBA system generalizes
the equations (\ref{wu-eq}) for abelian 
exclusion statistics. While many aspects of this work rely on
results of earlier work by various groups, several observations
are new. We demonstrated that the TBA system (\ref{order-k-eq})
has its origin in a non-diagonal, purely statistical $S$-matrix, 
which we explicitly specified for the case of $SU(2)_2$ spinons.
On the basis of the TBA system (\ref{order-k-eq}), we have
recovered and unified a number of results in the literature,
and we have been able to derive the defining thermodynamic
equations for quasi-holes over a new class of non-abelian 
quantum Hall states. We also illustrated the correspondence
between fermionic sum formulas and fractional statistics 
assignments in the case of non-abelian statistics.

\vskip 5mm

\noindent
{\bf Acknowledgments.} 
We have been informed that P.~Bouwknegt and L.-H.~Chim have made
observations that are very similar to those reported in this
paper. We thank them for communicating to us some of
their unpublished results \cite{BC}.  We thank P. Bouwknegt and O.~Warnaar
for
helpful suggestions. This work is supported in part by the 
foundation FOM of the Netherlands.

\end{document}